\documentclass[runningheads]{llncs}
\usepackage[T1]{fontenc}
\usepackage{graphicx}
\usepackage{color}
\usepackage{booktabs}
\usepackage{amsmath}
\usepackage{amssymb}
\usepackage[misc]{ifsym}
\usepackage{hyperref}
\usepackage{color}

\newcommand{\eg}{\textit{e.g.}}
\newcommand{\ie}{\textit{i.e.}}
\newcommand{\slope}{a_{\text{GA},\text{ID}}}

\begin{document}
%
\title{The Intrinsic Manifolds of Radiological Images and their Role in Deep Learning}
\titlerunning{The Intrinsic Manifolds of Radiological Images}
%
\author{Nicholas Konz\inst{1}\textsuperscript{*(\Letter)}\orcidID{0000-0003-0230-1598} \and Hanxue Gu\inst{1}\orcidID{0000-0003-2622-753X} \and Haoyu Dong\inst{2}\orcidID{0000-0002-5132-0341} \and Maciej A. Mazurowski\inst{1,2,3,4}\textsuperscript{(\Letter)}\orcidID{0000-0003-4202-8602} }

\authorrunning{N. Konz et al.}
%
\institute{Department of Electrical and Computer Engineering, Duke University, NC, USA \\ *\textit{corresponding author} \email{nicholas.konz@duke.edu}\and
Department of Radiology, Duke University, NC, USA \and
Department of Computer Science, Duke University, NC, USA \and
Department of Biostatistics \& Bioinformatics, Duke University, NC, USA \email{maciej.mazurowski@duke.edu}
}
\maketitle              
\begin{abstract}


The manifold hypothesis is a core mechanism behind the success of deep learning, so understanding the intrinsic manifold structure of image data is central to studying how neural networks learn from the data. Intrinsic dataset manifolds and their relationship to learning difficulty have recently begun to be studied for the common domain of natural images, but little such research has been attempted for radiological images. We address this here. First, we compare the intrinsic manifold dimensionality of radiological and natural images. We also investigate the relationship between intrinsic dimensionality and generalization ability over a wide range of datasets. Our analysis shows that natural image datasets generally have a higher number of intrinsic dimensions than radiological images. However, the relationship between generalization ability and intrinsic dimensionality is much stronger for medical images, which could be explained as radiological images having intrinsic features that are more difficult to learn. These results give a more principled underpinning for the intuition that radiological images can be more challenging to apply deep learning to than natural image datasets common to machine learning research.  We believe rather than directly applying models developed for natural images to the radiological imaging domain, more care should be taken to developing architectures and algorithms that are more tailored to the specific characteristics of this domain. The research shown in our paper, demonstrating these characteristics and the differences from natural images, is an important first step in this direction.

\end{abstract}

\section{Introduction}

Although using deep learning-based methods to solve medical imaging tasks has become common practice, there lacks a strong theoretical understanding and analysis of the effectiveness of such methods. This could be a potential problem for future algorithm development, as most successful methods for medical images are adapted from techniques solving tasks using natural image datasets \cite{chrabaszcz2017downsampled}.
Due to the apparent differences in relevant semantics between natural and medical domains \cite{morra2021bridging}, it is not clear what design choices are necessary when adapting these networks to medical images. This difference in domain is especially true when considering radiological images. Our goal is to provide a better, quantified footing for developing such radiology-specialized methods, by (1) analyzing the underlying structure of common radiological image datasets and determining how it relates to learning dynamics and generalization ability and (2) comparing these characteristics to common natural image datasets.

\begin{figure}
\centering
\includegraphics[width=0.95\textwidth]{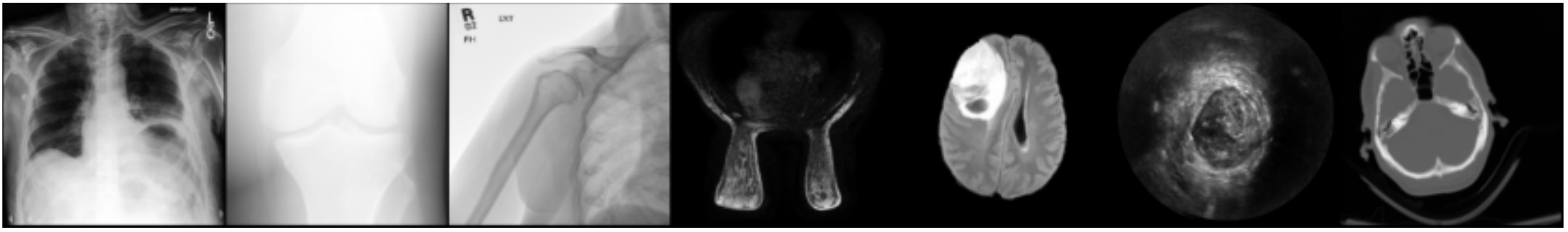}
\caption{\textbf{Sample images from each dataset studied.} From the left: CheXpert \cite{irvin2019chexpert}, OAI \cite{nevitt2006osteoarthritis}, MURA \cite{rajpurkar2017mura}, DBC \cite{saha2018machinedukedbc}, BraTS 2018 \cite{menze2014multimodal}, Prostate-MRI \cite{sonn2013prostate} and RSNA-IH-CT \cite{flanders2020rsnaihct} (see Sec. \ref{sec:data}).}
\label{fig:data_eg}
\end{figure}


The Manifold Hypothesis \cite{fefferman2016manifoldhyp,tenenbaum2000global,brand2002charting} states that high-dimensional data, such as images, can be well described by a much smaller number of features/degrees of freedom than the number of pixels in an image; this number is the \textit{intrinsic dimension} (ID) of the dataset.
This is central to deep computer vision because these abstract visual features can be learned from data, allowing inference in a tractable, lower-dimensional space. It is therefore important to study the relationship of the ID of datasets with the learning process of deep models. This was recently explored for standard natural image datasets \cite{pope2021intrinsic}, but a similarly comprehensive study has yet to be conducted for radiological image datasets, which is important because of the apparent differences between these two domains. 
\subsubsection{Contributions.}
\begin{figure}
\centering
\includegraphics[width=0.95\textwidth]{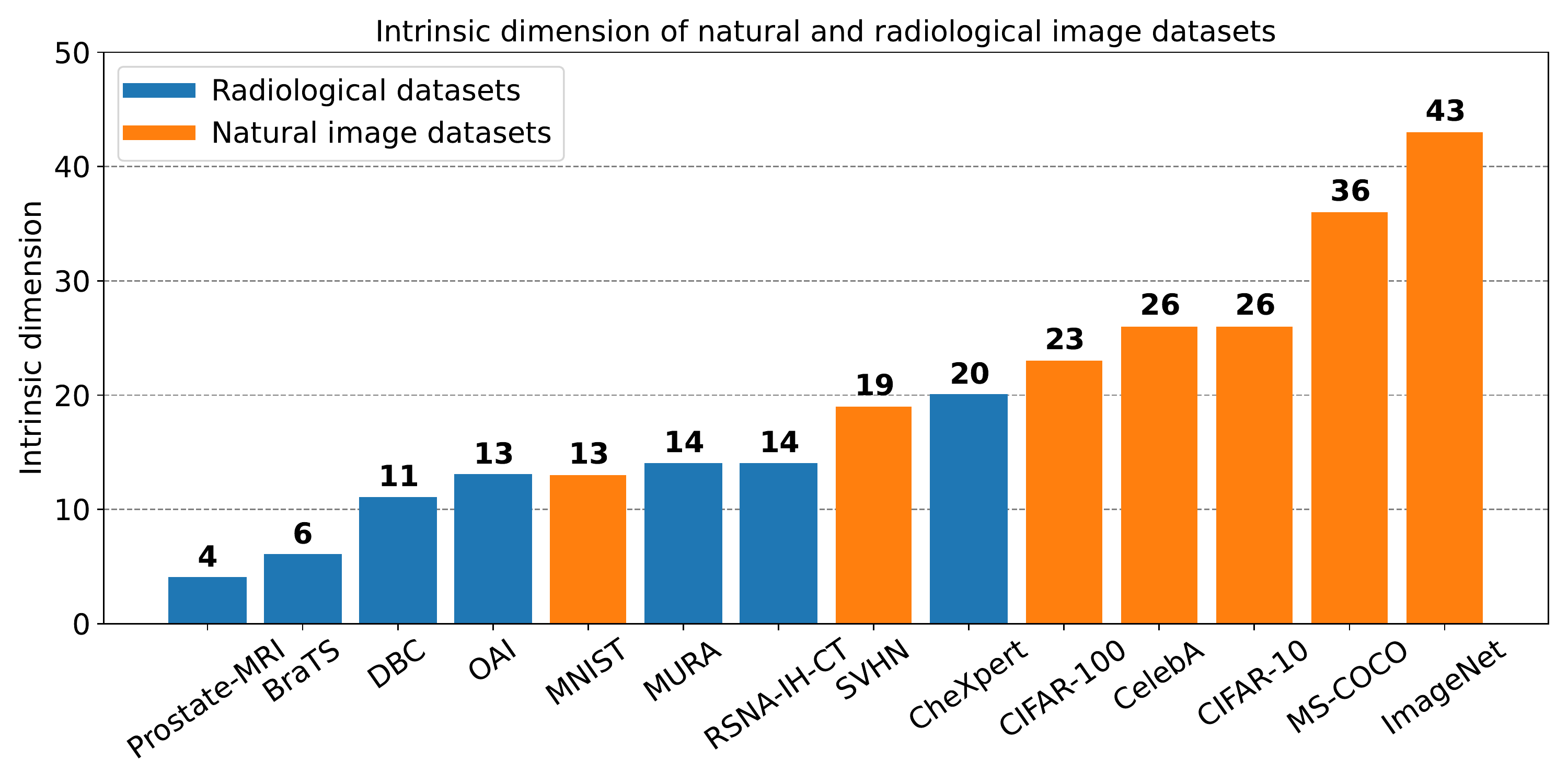}
\caption{\textbf{Intrinsic dimension of radiological (blue) and natural image (orange) datasets}, the latter from \cite{pope2021intrinsic}. Figure recommended to be viewed in color.} 
\label{fig:ID}
\end{figure}

Our contributions are summarized as follows.
\begin{enumerate}
    \item We investigate the intrinsic manifold structure of common radiology datasets (Fig. \ref{fig:data_eg}), and find that their IDs are indeed much lower than the number of pixels, and also generally lower than for natural image datasets (Fig. \ref{fig:ID}).
    \item We also find that classification is generally harder with radiology datasets than natural images for moderate-to-low training set sizes.  
    \item We show that classification performance is negatively linear to dataset ID within both data domains, invariant to training set size. However, the absolute value of the slope of this relationship is much higher for radiological data than for natural image data.
    \item We test these linearity findings on a wide range of common classification models, and find that performance for radiological images is almost independent of the choice of model, relying instead on the ID of the dataset.
\end{enumerate}


Our results show that while the ID of a dataset affects the difficulty of learning from this dataset, what also matters is the complexity of the intrinsic features themselves, which we find to be indicated by the sharpness of the relationship between generalization ability and ID, that is more severe for radiological images. These findings give experimental evidence for the differences in intrinsic dataset structure and learning difficulty between the two domains, which we believe is the first step towards a more principled foundation for developing deep methods specially designed for radiology.

\subsubsection{Related Work.}

Intrinsic dimension (ID) estimation methods (\cite{bac2021scikit}), have only recently been applied to datasets used in modern computer vision, beginning with \cite{pope2021intrinsic}, which explored the ID of common natural image datasets and how it relates to learning ability and generalization. There have been a few studies of the ID of medical datasets, \eg, \cite{cordes2006estimation}, but these are targeted at an individual modality or dataset. The most common ID estimator obtains a maximum-likelihood (MLE) solution for the ID by modeling the dataset as being sampled from a locally uniform Poisson process on the intrinsic data manifold \cite{levina2004maximum}. Other estimators exist (\cite{gomtsyan2019geomle,facco2017twonn}), but these are unreliable estimators for images \cite{pope2021intrinsic}, so we use the MLE estimator in this work.
Note that we do not estimate dataset ID with some learned latent space dimension found by training an autoencoder-type model or similar; relatedly, we also do not study the ID of the learned feature structure, or parameters of a trained feature extraction model, \eg, \cite{ansuini2019intrinsic,birdal2021intrinsic}. In contrast, we study the intrinsic, model-independent structure of the dataset itself.

\section{Methods: Intrinsic Dimension Estimation}
\label{sec:ID_est}

Consider some dataset $\mathcal{D}\subset\mathbb{R}^d$ of $N$ images, where $d$ is the \textit{extrinsic dimension}, \ie, the number of pixels in an image. The Manifold Hypothesis \cite{fefferman2016manifoldhyp} assumes that the datapoints $x\in\mathcal{D}$ lie near some low-dimensional manifold $\mathcal{M}\subseteq\mathbb{R}^d$ that can be described by an intrinsic dimension $\mathrm{dim}(\mathcal{M})= m\ll d$.

In order to obtain an estimate of $m$, \cite{levina2004maximum} models the data as being sampled from a Poisson process within some $m$-dimensional sphere at each datapoint $x$; the density of points is assumed to be approximately constant within the radius of the sphere. Rather than specifying the radius as a hyperparameter, the authors set this radius to be the distance of the $k^{th}$ nearest neighboring datapoint to $x$ (instead specifying $k$). By maximizing the likelihood of the data given the parameters of the Poisson model, we then obtain an ID for $x$ of
\begin{equation}
    \hat{m}_k(x) = \left[\frac{1}{k-1} \sum_{j=1}^{k-1} \log \frac{T_{k}(x)}{T_{j}(x)}\right]^{-1},
\end{equation}
where $T_j(x)$ is the $\ell_2$ distance from $x$ to its $j^{th}$ nearest neighbor. The authors of \cite{mackay2005comments} then showed that MLE can be used again to obtain an estimate for the global dataset ID, as
\begin{equation}
\label{eq:id_mle}
    \hat{m}_{k}=\left[\frac{1}{N} \sum_{i=1}^{n} \hat{m}_{k}\left(x_{i}\right)^{-1}\right]^{-1}=\left[\frac{1}{N(k-1)} \sum_{i=1}^{N} \sum_{j=1}^{k-1} \log \frac{T_{k}\left(x_{i}\right)}{T_{j}\left(x_{i}\right)}\right]^{-1},
\end{equation}
which is the estimator that we use for this work. Note here that $k$ is a hyperparameter of the estimator; just as in \cite{pope2021intrinsic} we set $k=20$, a moderate value as recommended by \cite{levina2004maximum}.
\section{Datasets and Tasks}
\label{sec:data}

In this work, we use the common task of binary classification to analyze the effect of dataset intrinsic dimension on the learning of radiological images. We chose radiology datasets that are varied in modality and well-representative of the domain, while being large enough for a broad study of training set sizes for at least one realistic classification task. We explore using alternate tasks for the same datasets in Section \ref{sec:ablation_task}. 

The datasets are as follows. \textbf{(1)} CheXpert \cite{irvin2019chexpert}, where we detect pleural effusion in chest X-ray images.
Next is \textbf{(2)} the Knee Osteoarthritis Initiative (OAI) \cite{nevitt2006osteoarthritis}, where we select the OAI-released screening packages 0.C.2 and 0.E.1 containing knee X-ray images. Following \cite{antony2017automatic,Tiulpin2018}, we build a binary osteoarthritis (OA) detection dataset by combining Kellgren-Lawrence (KL) scores of $\{0,1\}$ as OA-negative and combining scores of  $\{2,3,4\}$ as positive.
\textbf{(3)} is MURA \cite{rajpurkar2017mura}, where we detect abnormalities in musculoskeletal X-ray images.
Next, \textbf{(4)} is the Duke Breast Cancer MRI (DBC) dataset \cite{saha2018machinedukedbc}, where we detect cancer in fat-saturated breast MRI volume slices; we take slices containing a tumor bounding box to be positive, and all other slices at least five slices away from the positives to be negative. 
We follow this same slice-labeling procedure for dataset \textbf{(5)}, BraTS 2018 \cite{menze2014multimodal} where we detect gliomas in T2 FLAIR brain MRI slices. 
Dataset \textbf{(6)} is Prostate-MRI-Biopsy \cite{sonn2013prostate}, where we take slices from the middle 50\% of each MRI volume, and label each slice according to the volume's assigned prostate cancer risk score; scores of $\{0, 1\}$ are negative, and scores of $\geq 2$, which correlates with risk of cancer \cite{prostateTCIA}, are positive.
Our final dataset \textbf{(7)} is the RSNA 2019 Intracranial Hemorrhage Brain CT Challenge (RSNA-IH-CT) \cite{flanders2020rsnaihct}, where we detect any type of hemorrhage in Brain CT scans. Sample images are shown in Fig. \ref{fig:data_eg}.


\section{Experiments and Results}
\subsection{The Intrinsic Dimension of Radiology Datasets}

We first estimate the intrinsic dimension (ID) of the considered radiology datasets; the results are shown in Figure \ref{fig:ID}, alongside the natural image dataset results of \cite{pope2021intrinsic}. For each dataset, we estimate the MLE ID (Equation \eqref{eq:id_mle}) on a sample of $7500$ images such that there is an exact $50/50$ split of negative and positive cases in the sample, to minimize estimator bias (although we found that when using fewer images, the estimates were little affected).
Like natural images, we found the ID of radiological images to be many times smaller than the extrinsic dimension; however, radiological image datasets tend to have lower ID than natural images datasets. Intuitively, we found that modifying the dataset extrinsic dimension (resizing the images) had little effect on the ID.


\subsection{Generalization Ability, Learning Difficulty and Intrinsic Dimension}
\label{sec:exp:main}
We now wish to determine what role the intrinsic dimension (ID) of a radiology dataset has in the degree of difficulty for a deep model to learn from it, and how this compares to the natural image domain. As in \cite{pope2021intrinsic}, we use the test set accuracy obtained when the model has maximally fit to the training set as a proxy for the generalization ability (GA) of the model on the dataset. We train and test with each dataset separately on it's respective binary classification task (Sec. \ref{sec:data}), for a range of models and training set sizes $N_\text{train}$, from $500$ to $2000$. We train for 100 epochs with the same hyperparameters for all experiments; further details are provided in the supplementary materials.
For each experiment with a studied dataset, we sample $2750$ images from the given dataset such that there is an exact $50/50$ split of negative and positive images. From this, we sample $750$ test images, and from the remaining $2000$ images, we sample some $N_{\text{train}}$ training examples.

\begin{figure}
\centering
\includegraphics[width=0.65\textwidth]{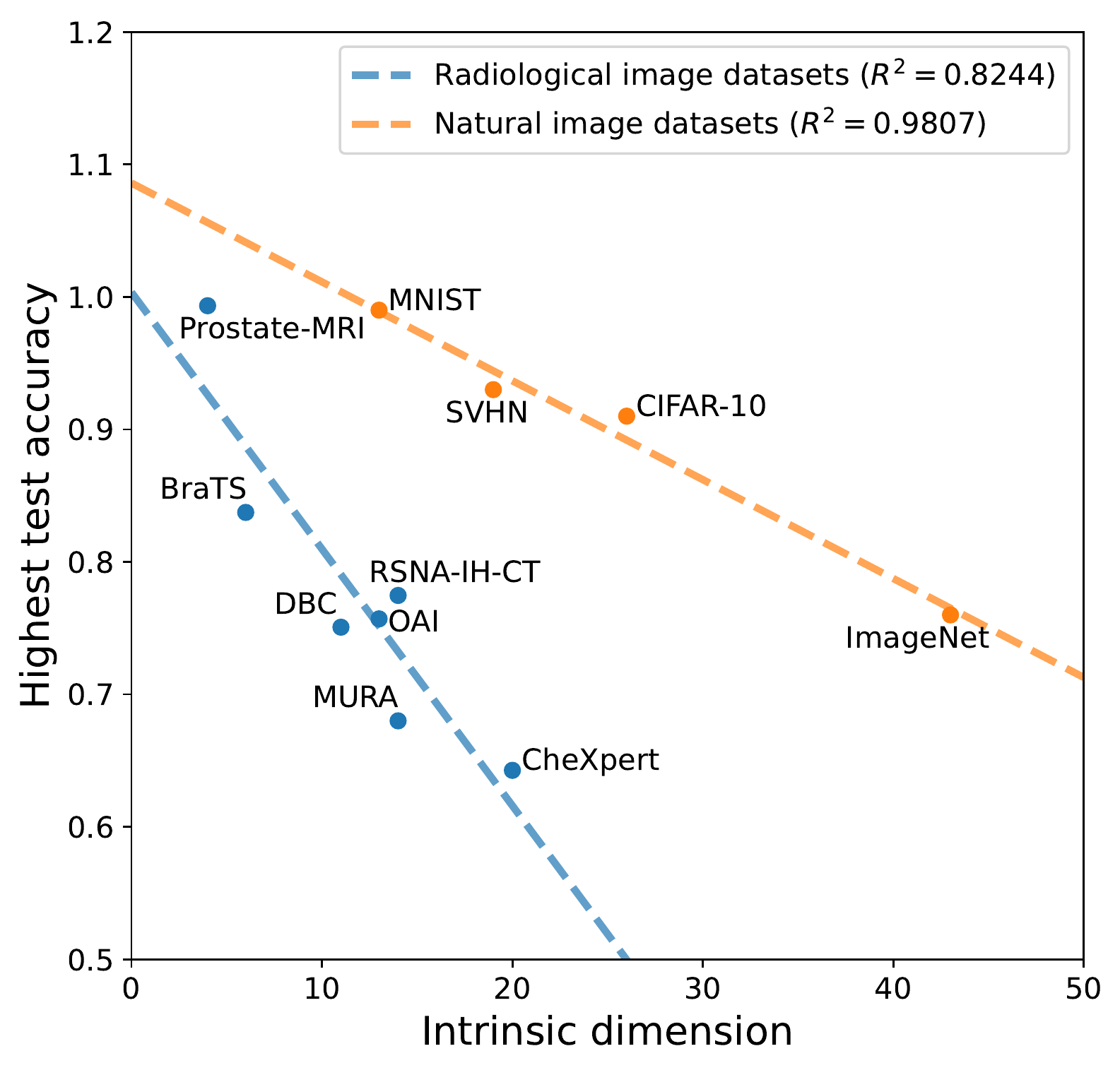}
\caption{\textbf{Linearity of model generalization ability with respect to dataset intrinsic dimension}, for radiological (blue) and natural (orange) image datasets, for $N_\text{train}=2000$ on ResNet-18. Figure recommended to be viewed in color.} 
\label{fig:multi_0}
\end{figure}

Beginning with $N_{\text{train}}=2000$ on ResNet-18, we plot the GA with respect to the dataset ID in Fig. \ref{fig:multi_0}, alongside the corresponding results for natural image datasets where binary classification was explored from \cite{pope2021intrinsic}. Intriguingly, even across the range of tasks and datasets, we see that the relationship of GA with ID is approximately linear within each domain. Indeed, when fitting a simple ordinary least squares linear model
\begin{equation}
    \label{eq:linregress}
    \text{GA} = a_{\text{GA},\text{ID}}\times\text{ID} + b
\end{equation}
to each of the two domains, we obtain a Pearson linear correlation coefficient of $R^2=0.824$ and slope of $\slope=-0.019$ for radiological images, and $R^2=0.981$ with $\slope=-0.0075$ for natural images. 

When repeating the same experiments over the aforementioned range of $N_\text{train}$ and ResNet models, we find that averaged over $N_\text{train}$, $R^2=0.76\pm0.05$, $\slope=-0.0199\pm0.0009$ and $R^2=0.91\pm0.12$, $\slope=-0.0077\pm0.0004$ ($\pm$ standard deviation) for the radiological and natural domains, respectively. These low deviations imply that even across a range of training sizes, both domains have a high, and mostly constant, correlation between GA and ID. Similarly, both domains have approximately constant slopes of GA vs. ID with respect to $N_\text{train}$, but between domains, the slopes differ noticeably.

\begin{figure}
\centering
\includegraphics[width=\textwidth]{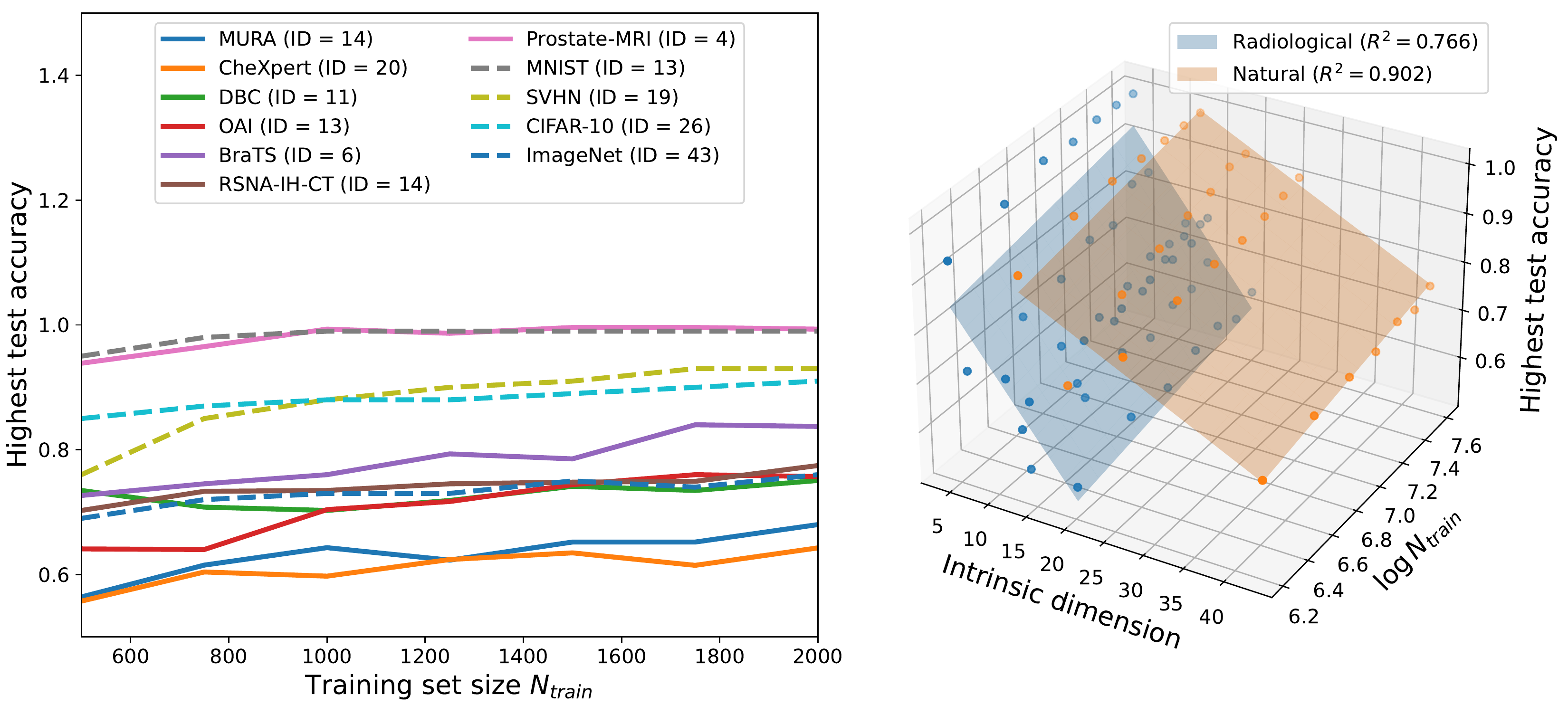}
\caption{\textbf{Left:} model generalization ability (GA) vs. training set size $N_\text{train}$ for various radiological (solid line) and natural (dashed line) image datasets, using ResNet-18. \textbf{Right:} linearity of GA with respect to $\log N_\text{train}$ and dataset intrisic dimension. Accompanies Fig. \ref{fig:multi_0}; recommended to be viewed in color.} 
\label{fig:multi_1}
\end{figure}

On the left of Fig. \ref{fig:multi_1}, we show how GA varies with respect to $N_\text{train}$ for datasets of both domains. We see that datasets with higher ID pose a more difficult classification task within both domains, \ie, more training samples are required to achieve some test accuracy. However, between these two domains, radiological images are generally more difficult to generalize to than natural images, for these moderate-to-low training set sizes (that are typical for radiology tasks \cite{soffer2019radiologydata2}). For example, OAI ($\text{ID}=13$), MURA ($\text{ID}=14$) and CheXpert ($\text{ID}=20$), all prove to be more difficult than ImageNet, even though it has more than double the number of intrinsic dimensions ($43$). 

This implies that the intrinsic dataset features described by these dimensions (and the correlations between them) can vary in learning difficulty between domains, indicated by the aforementioned sharper slope $\slope$. Our results show that radiological images generally have more difficult intrinsic features to learn than natural images, even if the number of these intrinsic feature dimensions is higher for the latter.

We can also explore the dependence of generalization ability (GA) on both training set size $N_\text{train}$ and ID, shown on the right of Fig. \ref{fig:multi_1}. \cite{narayanan2010sampleintrinsic} found that learning requires a training sample size that grows exponentially with the data manifold's ID; this implies that GA should scale with $\log N_\text{train}$. Indeed we see that GA is approximately linear with respect to ID and $\log N_\text{train}$: for each domain of radiological and natural, we find multiple linear correlation coefficients $R^2$ of $0.766$ and $0.902$ between these variables, respectively. Given some new radiology dataset, we could potentially use this to estimate the minimum $N_\text{train}$ needed to obtain a desired GA/test accuracy for the dataset, which would save overall training time. However, extensive experiments would need to be conducted before this could be widely applicable, due to confounding factors such as if the chosen model has a high enough capacity to fully fit to the training set; as such, we leave such an investigation for future work.

\subsubsection{Dependence on Model Choice.}

As mentioned, we repeated the same experiments with a number of additional models for the radiology domain, to see if these linear relationships change with different model choices. In addition to ResNet-18, we tested on ResNet-34 and -50 (\cite{he2016resnet}), VGG-13, -16 and -19 (\cite{simonyan2014vgg}), Squeezenet 1.1 (\cite{iandola2016squeezenet}), and DenseNet-121 and -169 (\cite{huang2017densenet}). Averaged over $N_\text{train}$ and all models, we obtained $R^2=0.699 \pm 0.080$ and $\slope = -0.019 \pm 0.001$ (individual model results are provided in the supplementary materials). By the low deviations of both $R^2$ and the actual regressed slope $\slope$ of GA vs. ID, we see that the same linear relationship between GA and ID also exists for these models. We therefore infer that this relationship between classification GA and ID is largely independent of model size/choice, assuming that the model has a high enough capacity to fully fit to the training set.

\subsubsection{Dependence on Task Choice.}
\label{sec:ablation_task}

Logically, there should be other factors affecting a model's GA beyond the dataset's ID; \eg, for some fixed dataset, harder tasks should be more difficult to generalize to. 
This section aims to determine how changing the chosen tasks for each dataset affects the preceding results. We will consider realistic binary classification tasks that have enough examples to follow the dataset generation procedure of Sec. \ref{sec:exp:main}. The datasets that support this are CheXpert---detect edema instead of pleural effusion, RSNA-IH-CT---detect subarachnoid hemorrhage, rather than any hemorrhage, and Prostate-MRI---detect severe cancer ($\text{score}>2$), rather than any cancer. For robustness we will experiment on all three aforementioned ResNet models.

From switching all three datasets to their modified tasks, the linear fit parameters (averaged over $N_\text{train}$) changed as $R^2=0.76\pm0.05\Rightarrow 0.78\pm0.15$ and $\slope=-0.0199\pm 0.0009\Rightarrow -0.012\pm 0.002$ for ResNet-18; $R^2=0.77\pm0.07\Rightarrow 0.76\pm0.21$ and $\slope=-0.019\pm 0.001\Rightarrow -0.011\pm0.003$ for ResNet-34; and $R^2=0.78\pm0.07\Rightarrow 0.82\pm0.15$ and $\slope=-0.021\pm 0.001\Rightarrow -0.013\pm 0.004$ for ResNet-50. We therefore conclude that the choice of task has some, but a small effect on the significance of the linear relationship of GA with ID $(R^2)$, but can affect the parameters (slope $\slope$) of the relationship. Certainly, the choice of task represents a very large space of possibilities, so we leave more comprehensive investigations for future work.

\section{Conclusion} 

Our results provide empirical evidence for the practical differences between the two domains of radiological and natural images, in both the intrinsic structure of datasets, and the difficulty of learning from them. We found that radiological images generally have a lower intrinsic dimension (ID) than natural images (Fig. \ref{fig:ID}), but at the same time, they are generally harder to learn from (Fig. \ref{fig:multi_1}, left). This indicates that the intrinsic features of radiological datasets are more complex than those in natural image data. Therefore, assumptions about natural images and models designed for natural images should not necessarily be extended to radiological datasets without consideration for these differences. Further study of the differences between these two domains and the conceptual reasons for why they arise could lead to helpful guidelines for deep learning with radiology. We believe that the results in this work are an important step in this direction, and they lay the foundation for further research on this topic.


{\small
\bibliographystyle{plain}
\bibliography{paper1636}
}

\newpage
\section*{Supplementary Materials for ``The Intrinsic Manifolds of Radiological Images and their Role in Deep Learning''}
\appendix 

\section{Experimental Settings (for reproducibility)}

\begin{table}
\centering
\caption{Training/Inference Parameters and Settings}\label{tab:app:hyper}
    \begin{tabular}{|l|l|}
        \hline
        Name & Value \\
        \hline
        Loss Function & Binary Cross-Entropy \\
        Optimizer & Adam \cite{kingma2014adam} \\
        Learning rate & 0.001 \\
        Weight decay & 0.0001 \\
        Image normalization & $[0, 255]$ \\
        Image resolution & $224\times 224$ \\
        Training image augmentations & None \\
        GPUs & $4\times$ NVIDIA GeForce RTX 3070 (8 GB)
        \\
        Experiment Software & PyTorch 1.8.1, Numpy 1.22.1, CUDA 11.4 \\
        Visualization Software & MatPlotLib 3.4.2 \\
        Typical model training time (on $N_\text{train}=2000$) & 10-30 min. \\
        
        \hline
    \end{tabular}

\end{table}

\begin{table}
\centering
\caption{Model-Specific Details and Results (averaged over $N_\text{train}$ with std. dev.)}\label{tab:app:models}
    \begin{tabular}{|l|c|c|c|}
        \hline
        Model & Training batch size & $R^2$ (GA vs. ID) & slope $\slope$ (GA vs. ID)\\
        \hline
        ResNet-18 \cite{he2016resnet} & 200 & $0.756 \pm 0.052$ & $-0.0199 \pm 0.0009$ \\
        ResNet-34 \cite{he2016resnet} & 128 & $0.772 \pm 0.071$ & $-0.0193 \pm 0.0012$ \\
        ResNet-50 \cite{he2016resnet} & 64 & $0.781 \pm 0.066$ & $-0.0207 \pm 0.0010$ \\
        VGG-13 \cite{simonyan2014vgg} & 32 & $0.646 \pm 0.048$ & $-0.0194 \pm 0.0009$ \\
        VGG-16 \cite{simonyan2014vgg} & 32 & $0.623 \pm 0.066$ & $-0.0184 \pm 0.0008$ \\
        VGG-19 \cite{simonyan2014vgg} & 32 & $0.597 \pm 0.100$ & $-0.0168 \pm 0.0031$ \\
        Squeezenet 1.1 \cite{iandola2016squeezenet} & 32 & $ 0.580 \pm 0.073$ & $-0.0173 \pm 0.0011 $ \\
        DenseNet-121 \cite{huang2017densenet} & 32 & $ 0.770 \pm 0.073 $ & $ -0.0190 \pm 0.0009$ \\
        DenseNet-169 \cite{huang2017densenet} & 32 & $ 0.765 \pm 0.061$ & $-0.0189 \pm 0.0008$ \\
        \hline
    \end{tabular}
\end{table}

\end{document}